\documentclass[aps,prb,twocolumn,superscriptaddress,showpacs]{revtex4-1}

\usepackage{hyperref}
\usepackage[dvips]{graphicx}
\usepackage{amssymb,amsmath,amsfonts}
\usepackage[T1]{fontenc}
\usepackage{float}
\usepackage{color,soul}
\usepackage{booktabs}
\usepackage{color}
\usepackage{tabularx}
\usepackage{physics}

\begin{document}

\title{Quantum computation of silicon electronic band structure}

\author{Frank T. Cerasoli}
  \affiliation{Department of Physics, University of North Texas, Denton, TX 76203, USA}

\author{Kyle Sherbert}
 \affiliation{Department of Physics, University of North Texas, Denton, TX 76203, USA}

\author{Jagoda S\l awi\'{n}ska}
\affiliation{Department of Physics, University of North Texas, Denton, TX 76203, USA}

\author{Marco \surname{Buongiorno Nardelli}}
\email{Email: mbn@unt.edu}
\affiliation{Department of Physics, University of North Texas, Denton, TX 76203, USA}


\begin{abstract}
Development of quantum architectures during the last decade has inspired hybrid classical-quantum algorithms in physics and quantum chemistry that promise simulations of fermionic systems beyond the capability of modern classical computers, even before the era of quantum computing fully arrives. Strong research efforts have been recently made to obtain minimal depth quantum circuits which could accurately represent chemical systems. Here, we show that unprecedented methods used in quantum chemistry, designed to simulate molecules on quantum processors, can be extended to calculate properties of periodic solids. In particular, we present minimal depth circuits implementing the variational quantum eigensolver algorithm and successfully use it to compute the band structure of silicon on a quantum machine for the first time. We are convinced that the presented quantum experiments performed on cloud-based platforms will stimulate more intense studies towards scalable electronic structure computation of advanced quantum materials.

\end{abstract}
\maketitle

\section{Introduction}
Quantum computing aims to leverage superposition, entanglement and interference of quantum bits in order to tackle computational tasks that scale exponentially on classical computers.\cite{nielsen, Abrams_1999} While renowned quantum algorithms, such as unsorted database search or integer factorization require resources that remain out of reach,\cite{shor, grover} quantum chemistry calculations are gaining steam as a key application performed on available quantum architectures.\cite{review_mod, review_chem} The idea of so-called quantum simulations originally proposed by Feynman,\cite{feynman, lloyd, entropy} relies on a mapping between the fermionic system and the set of qubits, so that the dynamics of the former is directly followed by the latter. Therefore, wave functions of complex many-body systems could be effectively reproduced in quantum measurements performed on qubits, providing a tool to compute desired quantities with an unprecedented accuracy. Even though available quantum computers contain merely few tens of qubits,\cite{google} they have been employed to solve quantum chemistry problems, such as the estimation of nuclear binding energies or molecular ground states.\cite{deuteron, beh2, guzik_lih, water} Remarkably, these successful quantum experiments relied on variational approaches that greatly reduced the required hardware resources, inspiring more active research in order to solve elusive condensed matter systems beyond quantum chemistry. \cite{babbush, troyer_dmft, kreula_epj, rungger_dmft, gutzwiller}

Here, we put forward an approach to calculate the electronic structure of the periodic crystal on a quantum computer. While developments of quantum computation for molecules were primarily focused on the ground state energies, to evaluate a band structure one needs to determine the excited states. We have shown that a standard hybrid quantum/classical algorithm, variational quantum eigensolver (VQE) can be easily adapted to provide an accurate estimation of the electronic bands in the solid. In particular, by casting a Si tight-binding (TB) Hamiltonian in terms of fermionic operators, we have designed a low-depth quantum circuit, enough robust to capture the electronic properties of a crystal in the reciprocal space. The quantum measurements have been performed on sets of qubits available remotely via cloud-based platforms provided by IBM and Rigetti Computing. Importantly, we have tested different classical optimization routines that minimize expectation values, corrected beforehand against the readout errors. Comparison between bands computed on the quantum processors, the quantum virtual machine and by classical diagonalization revealed a satisfactory agreement, confirming validity of the algorithm which could be generalized to explore materials more complex than crystalline silicon.


\section{Hamiltonian Representation}
Let us consider a silicon lattice in the diamond cubic structure. The Hamiltonian describing the electronic system can be approximated, in atomic units, as

\begin{equation}\label{eq:tb_1}
\hat{H}_{el}=-\sum_{i}{\frac{\nabla^{2}_{\vb*{r}_{i}}}{2}}-\sum_{i,j}{\frac{Z_j}{|\vb*{R}_{j}-\vb*{r}_{i}|}}+\frac{1}{2}\sum_{i\neq l}{\frac{1}{|\vb*{r}_{l}-\vb*{r}_{i}|}}
\end{equation}

\noindent where $\vb*{r}_{i}$ ($\vb*{R}_{i}$) are the positions of electrons (nuclei) and $Z_j$ denotes the nuclear charge, respectively. We have assumed the Born-Oppenheimer approximation and considered the nuclei as stationary charges, thus neglecting their kinetic energy and treating the ion repulsion as a constant. The last term of Eq.\ref{eq:tb_1} represents the electron-electron interaction, whose correct estimation is one of the long-term goals of quantum simulation. However, we are now primarily focused on the proof-of-principle band structure calculations, and have disregarded the electronic correlations for the purpose of the present study.

In order to convert the Hamiltonian into a computational problem, a suitable basis set needs to be selected. While different representations were proposed for quantum computation,\cite{babbush} we introduce here a simple basis of atomic orbitals at each lattice site arising from the tight-binding (TB) approximation. The unit cell of silicon contains two tetrahedrally coordinated ions and is well described in terms of $s$, $p_x$, $p_y$ and $p_z$ orbitals centered at each atom. Because magnetic order is absent, the spin degrees of freedom can be omitted in the analysis. Using the second quantization formalism, we can express the TB Hamiltonian via creation and annihilation operators ($a^{\dagger}_{in}$ and $a_{in}$) acting at the orbital $n$ and the site $R_i$:

\begin{equation}\label{eq:tb_2}
\hat{H}=\sum_{i,n}E_{n}a^{\dagger}_{in}a_{in}-\sum_{<i,j>,n,m}t_{in,jm}a^{\dagger}_{in}a_{jm}
\end{equation}

\noindent In this expression, $E_{n}$ correspond to the atomic energies and $t_{in,jm}$ denote the hopping integrals whose numerical values have been reported elsewhere.\cite{Cohen_TB} Only the tunneling between pairs of nearest neighbors, denoted by the $<i,j>$ summation, have been considered. The Hamiltonian can be then easily converted to the momentum space via standard Fourier transform applied to the raising and lowering operators. Last, such a representation ($\hat{H}_{\vb*{k}}$) needs to be mapped onto the system of qubits.

In practice, qubits are manipulated on a quantum processor by operating on a set of Pauli matrices $X$, $Y$, $Z$ and $I$, the latter denoting $2\times2$ identity matrix. Any Hermitian matrix can be decomposed using a complete Pauli basis for matrices of dimension $N=2^n$ with $n=\lceil$log$_2 N\rceil$ terms, that can be generated by taking a tensor product:

\begin{equation}\label{eq:pauli_decomp}
\{\hat{\sigma}\}_n=\{I,X,Y,Z\}^{\otimes n}
\end{equation}
\newline
\noindent Thus, TB Hamiltonian can be decomposed as follows:

\begin{equation}\label{eq:tb_ham}
\hat{H}_{\vb*{k}}=\sum_{i=1}^{4^n}{c_{i\vb*{k}}\hat{\sigma}_i}
\end{equation}
\newline
\noindent where the set $\{\hat{\sigma}\}_{n}$ is the set of $4^n$ possible basis matrices, and $\{c_{\vb*{k}}\}_{n}$ is a set of complex coefficients. $\{c_{\vb*{k}}\}_{n}$ is known as the \textit{spectral decomposition} and can be  determined easily. In particular, we can exploit the orthogonality of Pauli matrices and the trace inner product between two of them:

\begin{equation}\label{eq:pauli_ortho}
\Tr(\hat{\sigma}_i^\dagger\hat{\sigma}_j)=2^n\delta_{ij}
\end{equation}

\noindent By taking the inner product $\Tr\lparen\hat{H_{\vb*{k}}}^\dagger\hat{\sigma}_i\rparen$, we can eliminate all terms but one from the sum, yielding:

\begin{equation}\label{eq:pauli_ck}
c_i=\frac{\Tr(\hat{H}_{\vb*{k}}^\dagger\hat{\sigma}_i)}{2^n}
\end{equation}

Therefore, the Hamiltonian is represented by a list of coefficients corresponding to each of the $4^n$ Pauli basis matrices suitable for simulation on a QPU.\cite{unitary_partitioning}

\section{Variational Quantum Eigensolver}
We have computed the energy spectrum using the variational quantum eigensolver in conjunction with overlap-based techniques. VQE is a standard hybrid quantum-classical algorithm capable to determine the lowest or highest eigenvalue of an operator using minimal quantum resources, implemented by combining measurements on a quantum computer with classical routines.\cite{Peruzzo2014,npj,prx} The ground state wave function and energy can be found based on Rayleigh-Ritz variational principle, whereby the energy expectation value can be minimized by a specific set of parameters. In practice, the state preparation and the expectation value measurements are implemented on a quantum machine, while the optimization of the parameters is performed classically. The whole algorithm used for the ground state calculation can be summarized in three following steps:

\begin{enumerate}
  \item We create a quantum circuit $\hat{V}(\vb*{\theta})$ depending on a set of parameters $\vb*{\theta}$, known as a \textit{variational form}. Then, we prepare a trial wave function (or \textit{ansatz}) $|\psi(\vb*{\theta})\rangle=\hat{V}|\vb*{0}\rangle$, where $|\vb*{0}\rangle$ denotes an initial state ensuring the measurement of each qubit.
  \item We measure the expectation value of $\hat{H}_{\vb*{k}}$, which depend on the parameters $\vb*{\theta}$, $E(\vb*{\theta})=\langle\psi(\vb*{\theta})|\hat{H}_{\vb*{k}}|\psi(\vb*{\theta})\rangle$. The Hamiltonian is represented by series of operators. The wave function $|\psi\rangle$ is measured in the Pauli basis, yielding each $\langle\sigma_i\rangle$. We can then reconstruct  $\langle\hat{H}_{\vb*{k}}\rangle$ with the spectrum $\{\vb*{c_k}\}$:
\begin{equation}\label{eq:ham_expectation}
\langle\hat{H}_{\vb*{k}}\rangle=\sum_{i=1}^{4^n}c_{i\vb*{k}} {\langle\hat{\sigma}_i}\rangle
\end{equation}

  The measurement should be treated as a probabilistic element of the algorithm and needs to be performed several times. An arbitrary precision can be achieved with a sufficient number of repetitions.

  \item We apply a classical optimization routine to explore the parameter space and minimize $E(\vb*{\theta})$. We define $\epsilon_0=\langle\psi(\vb*{\theta_{min}})|\hat{H}_{\vb*{k}}|\psi(\vb*{\theta_{min}})\rangle$ as a ground state energy, where $\vb*{\theta_{min}}$ denotes the set of parameters minimizing the expectation value of $\hat{H}_{\vb*{k}}$.

\end{enumerate}

\section{Energies Beyond the Ground State}
After having determined the ground state, we can calculate excited states using a procedure similar to the quantum deflation algorithm that exploits orthogonality of the Hamiltonian eigenvectors.\cite{Higgott2019variationalquantum, numerical_overlap, endo} In particular, we define an effective Hamiltonian ($\hat{H}'$) whose lowest eigenstate is the excited state of the original one ($\hat{H}$). By subtracting from the latter a corresponding ground state projector weighted by the ground state energy, we obtain:

\begin{equation}
\hat{H}_{\vb*{k}}'=\hat{H}_{\vb*{k}}-\epsilon_0|\psi_0\rangle\langle\psi_0|=\sum_{i=1}^{4^n}{(c_i-\epsilon_0 \frac{\langle \hat{\sigma_i} \rangle}{2^n}) \hat{\sigma_i}}
 \end{equation}

\noindent We observe that the last equality provides the following spectral decomposition of the excited Hamiltonian:
\begin{equation}
c'_i=c_i-\epsilon_0 \frac{\langle \hat{\sigma_i} \rangle}{2^n}
\end{equation}
The procedure is used iteratively to determine as many eigenvalues as desired. Updating the spectral decomposition $c_i \rightarrow c_i - \epsilon_0 \frac{\langle \hat{\sigma_i} \rangle}{2^n}$ effectively removes all ground state contributions from the Hamiltonian.

We note that the effect of subtracting the ground state density matrix weighted by its corresponding eigenvalue is to project that eigenstate onto the zero value. Because an arbitrary Hermitian matrix can have both positive and negative eigenvalues, special care must be taken to ensure that the zero is not erroneously computed as a ground state after all negative eigenvalues are determined. One reconciliation is to subtract a value greater than the maximum eigenvalue from the diagonal elements of the Hamiltonian, ensuring that all eigenvalues are lower than zero. Therefore, projecting an eigenstate to zero would not affect the remaining eigenvalues that need to be determined. Such a shift requires the modification of only one coefficient of the spectral decomposition, which stands before the identity matrix.

\section{Data Acquisition}
Before discussing the results of quantum experiments, let us remark  on the various techniques that we have employed to compute the band structure of silicon. A careful distinction must be made between the use of quantum processor, quantum virtual machine and quantum state simulation. In particular, simulated qubits helped us analyze the performance of variational forms and the effect of measurement uncertainty on a noiseless machine. Three independent techniques will be further referenced:

\begin{enumerate}
  \item \textit{Quantum Processor Unit} (QPU) is prepared for measurements under subsequent sets of parameters. The measurements are performed in real time. The available APIs compile quantum programs and directly manipulate qubits, providing measured expectation values in the form of bitstrings.
  \item \textit{Quantum Virtual Machine} (QVM) chooses one of the possible outcomes to be "measured", weighted by its respective probability computed with the quantum state simulator (see below). The quantum processor is mimicked, providing a noiseless (unless noise is simulated) simulation of the measurement process. This method helps to analyze the effects in the band structure determined by discrete measurements of the energy expectations values.
  \item \textit{Quantum State Simulator} (QSS) carries out linear algebra to obtain an exact wave function which would represent the simulated state of a qubit on a quantum processor after the application of specified gates. It can serve as an analytical guideline for quantum measurements. Optimization can be easily performed with the quantum state simulator, providing a convenient framework to test the performance of variational forms.
\end{enumerate}

\section{Quantum Experiments}

\begin{figure}[b]\label{fig:mf_circuit}
\includegraphics[scale=.145]{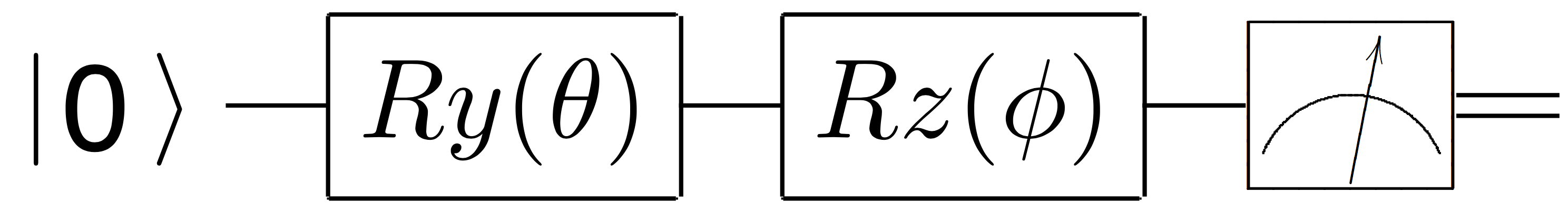}
\caption{Mean-field circuit acting on a single qubit has been employed to determine the lowest bands of silicon. It consists of a polar rotation ($R_y$) followed by an azimuthal rotation ($R_z$). In last step, the expectation value of $\hat{H}_{\vb*{k}}$ is measured.
}
\end{figure}

\begin{figure*}[ht]\label{fig:qpu_bands}
\includegraphics[scale=0.98]{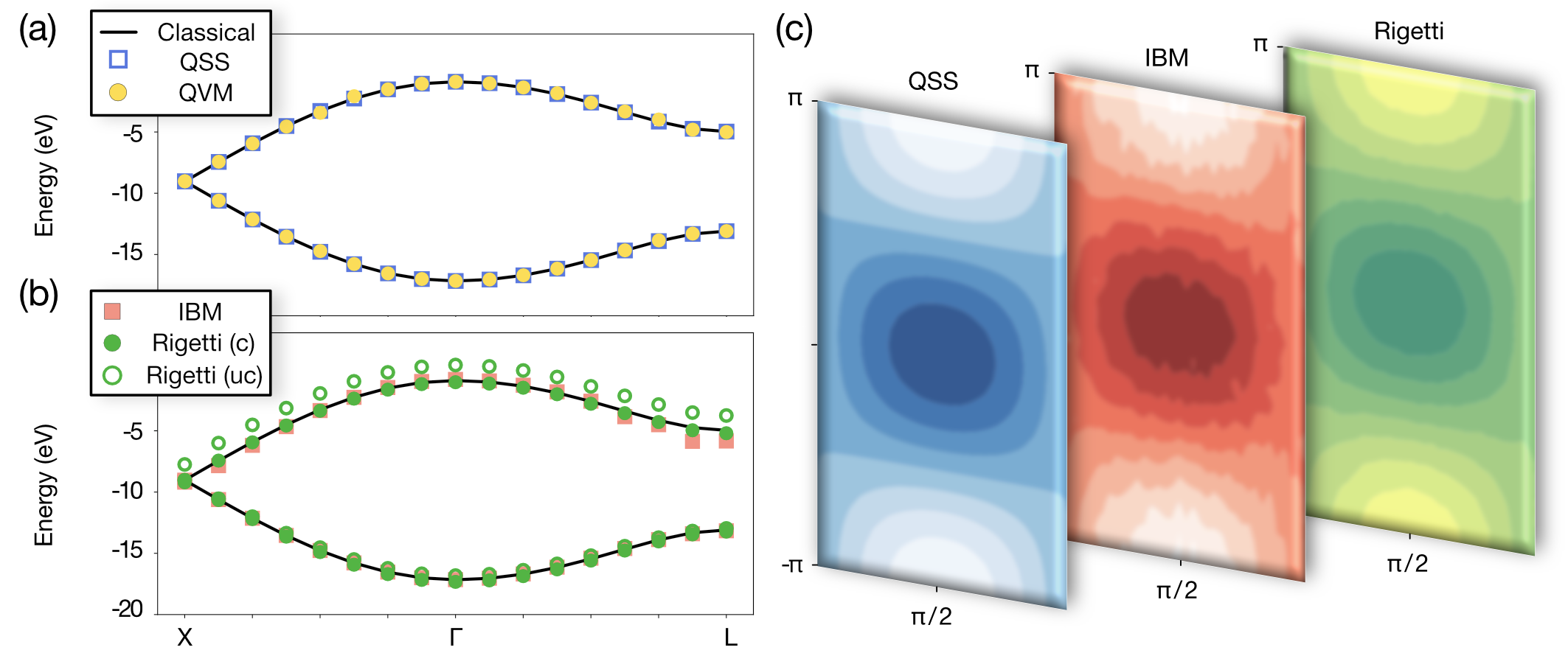}
\caption{(a) The two-band electronic structure of silicon computed along $X-\Gamma-L$ line using classical diagonalization (black solid line) and hybrid quantum-classical algorithm  on quantum state simulator (blue squares) and  quantum virtual machine (yellow circles). (b) Same as (a) realized on the QPUs of IBM (red squares) and Rigetti (green circles). We report the data from Rigetti before and after correcting for the readout errors, marked as open and closed circles, respectively. (c) Energy expectation value sampled over the entire parameter space $[-\pi,\pi]$ in the azimuthal angle and $[0,\pi]$ in the polar angle on QSS (blue), IBM (red) and Rigetti (red). Darker (brighter) colors denote lower (higher) values of the energy expectation value.}
\centering
\end{figure*}

Quantum computations of the band structure have been performed following two different techniques, both yielding a correct spectrum while compared with the classical diagonalization of the TB Hamiltonian. The first approach relies on a true quantum measurement, employing one qubit that we access on remote quantum machines Rigetti Aspen and IBMQ Armonk. Although these cloud platforms permit the use of larger resources, the practical realization of the VQE algorithm for diagonalization of the $8\times8$ Hamiltonian of Si required a substantial amount of time. Therefore, we have started with a reduced Hamiltonian, considering only the interactions between $s$-states which give rise to the lowest bands of silicon. After neglecting $s-p$ hopping parameters in the original $\hat{H}_{\vb*{k}}$, a smaller $2\times2$ matrix block can be decoupled and diagonalized using VQE on the QPU. Figure 1 shows the two-gate circuit acting on a single qubit, often referred to as the \textit{mean field} ansatz,~\cite{mean_field} which has been used in the experiment. In principle, to ensure that finding the true minimum is possible, circuits must be designed to span every state allowed by the operating qubits, unless the space is restricted by physical arguments, such as fermionic commutation relations in the UCC strategies.~\cite{Romero_2018} The ansatz below takes a pure state $|\vb*{0}\rangle$ and applies two rotations described by the angles $\vb*{\theta} = (\theta, \phi)$. A polar rotation brings the qubit into a superposition of $|\vb*{0}\rangle$ and $|\vb*{1}\rangle$ states, while an azimuthal rotation scans the sphere's latitude. The two rotations produce a state represented by the following wave function:

\begin{equation}
\ket{\psi(\theta,\phi)} = \cos(\frac{\theta}{2}) \ket{\vb*{0}} + e^{i\phi} \sin(\frac{\theta}{2}) \ket{\vb*{1}}
\end{equation}

The band structure has been computed along a high-symmetry line $X-\Gamma-L$ by repeating the whole algorithm for each of the $k$-points. Figures 2(a-b) report the two-band electronic structure evaluated on the quantum machines of IBM (red squares) and Rigetti (green circles), complemented by the data from the classical diagonalization (black solid line). In addition, we present the results obtained via quantum-classical algorithm performed on QSS (blue squares) and QVM (yellow circles). While the latter directly follow the bands calculated classically, the quantum data reveal tiny deviations that can be noticed around the high-symmetry points $\Gamma$ and $L$ for Rigetti and IBM, respectively. The sources of errors in the experiment can be manifold. The probabilistic aspect can obviously play a role, despite a large number of measurements (8192) taken for each parameterization. Importantly, simulation of noise on QVM have revealed that any gate noise or readout error tends to increase the measured energy, shifting the expectation value toward different eigenstates. As described in the next sections, we have attempted to characterize and reduce the effects of errors arising from the qubit manipulation.

\begin{figure*}[ht!]\label{fig:3q_circuit}
\includegraphics[scale=.29]{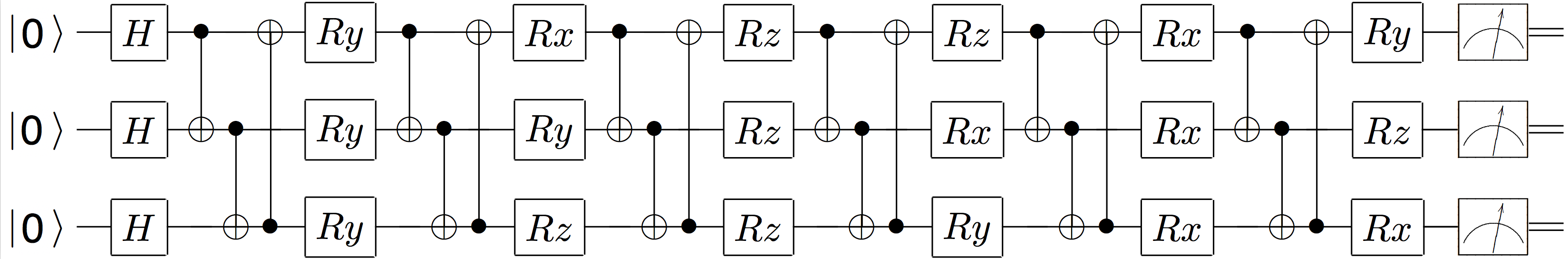}
\caption{The circuit used to diagonalize the $8\times8$ Hamiltonian. Each qubit is initialized as a pure zero state.}
\centering
\end{figure*}

We note that the standard optimization routines have not been here applied. Instead, we have used the mean-field circuit to measure a dense grid of parameter angles in order to find the minimum expectation value. Sampling the entire parameter space provides a visual tool for analyzing the structure of parameter space. Figure 2(c) shows examples of the expectation value surfaces computed for one selected point $\vb*{k}=\frac{\pi}{4a}\langle 1,1,1\rangle$. The three subsequent panels report the surfaces obtained analytically on QSS (blue) and experimentally on IBM (red) and Rigetti (green). The two latter have been smoothed by minimizing the root-mean-square error across all data points. Again, the data collected on IBM reveals largest irregularities in the energy contour lines, especially compared with the analytical surface evaluated on QSS.

The second approach, employed to diagonalize full $8\times8$ Hamiltonian, relies on QSS. Figure 3 presents a robust three-qubit circuit that we have designed to variationally minimize the expectation value of $\hat{H}_{\vb*{k}}$ at any $k$-point and each level of excitation. The set of twelve parameters $\vb*{\theta}=(\theta_1,\theta_2,...,\theta_{12})$ in this ansatz, measured in the Pauli word basis from the Hamiltonian decomposition defined in Eq.(\ref{eq:tb_ham}), are varied to minimize the energy expectation values. Figure (4) displays the electronic structure computed using this circuit, demonstrating that it is indeed capable of representing the silicon Hamiltonian anywhere along the $k$-line. Although small discrepancies are again visible, the overall agreement with the bands calculated classically seems to be sufficient. We note that now the results do not depend on external factors that can perturb the behavior of qubits. The deviations are related to the optimization procedures whose proper choice is essential to correctly determine the energy spectrum.

\begin{figure}[ht]
\includegraphics[scale=0.98]{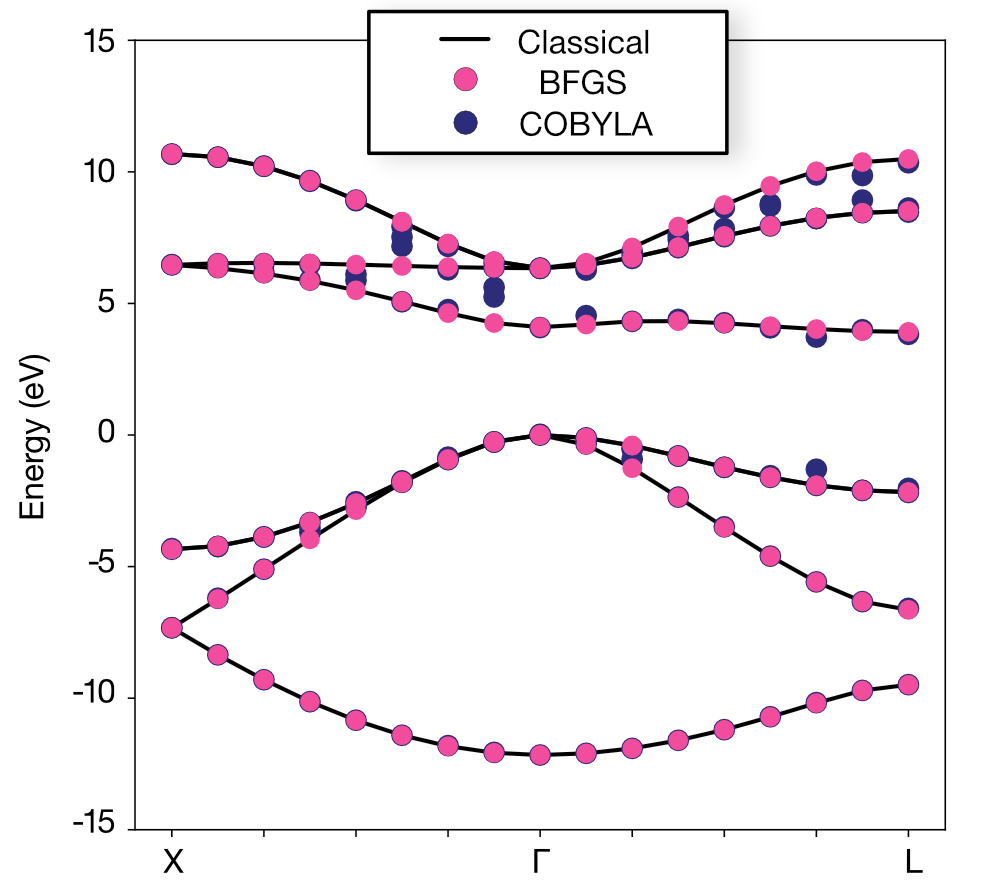}
\caption{Electronic structure of silicon computed via hybrid classical/quantum algorithm on QSS. Different optimization routines BFGS and COBYLA are compared on analytic surface. Black solid lines denote the bands calculated classically.}
\label{fig:sim_bands_8x8}
\end{figure}

Several classical optimization routines have been tested in conjunction with the three-qubit circuit used for the evaluation of full electronic structure. Minimizing a function in parameter space of twelve dimensions is rather challenging and requires a compromise between the number of measurements and the smoothness of the space being optimized. We have found that the Broyden-Fletcher-Goldfarb-Shanno (BFGS) and Constrained Optimization BY Linear Approximation (COBYLA) routines\cite{powell} yielded the most accurate results. The former requires fewer function evaluations to reach a minimum, but it suffers from instability due to the rough surface in parameter space. The latter, being a direct search method, entirely omits the idea of gradient decent which makes it more robust against becoming trapped in a local minimum. Even though it may provide more reliable global minima,\cite{Nelder1965ASM} it occasionally fails to settle on the correct set of parameters. Figure 4 clearly shows that especially the excited energy levels are sensitive to fluctuations in the determined parameters. The comparison of both routines, BFGS and COBYLA, eventually indicates the superior performance of the former, at least in the present case.

\section{Additional Remarks on Measuring Expectation Values}
While the previous section was entirely focused on the realization and results of quantum experiments, the measurements of expectation values need a more detailed discussion. The quantities we have measured on the quantum computer are the expectation values $\expval{\sigma_k}$, where the operator $\sigma_k$ is an $n$-length Pauli word consisting of an $I$, $X$, $Y$, or $Z$ for each qubit. They depend on the state $\ket{\psi}$ of the qubits, and could be written as the integral $\expval{\sigma_k}{\psi}$.
Because we do not know $\ket{\psi}$, we must measure the state of each qubit in the computational basis, resulting in a single bitstring (eg. $\ket{00101}$).
Repeating the measurement a large number of times $M$, we construct the expectation value $\expval{\sigma_k}$ from the ensemble of bitstrings.
In the following paragraphs, we will first consider the simple single-qubit case $\sigma_k=Z$, then the multi-qubit case where $\sigma_k$ consists only of $I$ and $Z$ operators and last, the general case including $X$ and $Y$ operators.

The Pauli operator $Z$ can be written in a matrix form:
$$ Z = \left[\begin{array}{rr} 1 & 0 \\ 0 & -1 \end{array}\right] $$
It is a diagonal matrix with eigenvalues $+1$, corresponding to the state $\ket{0}$, and $-1$, corresponding to the state $\ket{1}$.
The expectation value $\expval{Z}$ is the average of these two eigenvalues, weighted by the number of measurements in each state. If $p$ is the probability that we measure $\ket{0}$ rather than $\ket{1}$, the expectation value $\expval{Z}$ is given by:
$$ \expval{Z} = (+1)p + (-1)(1-p) = 2p - 1 $$

Now, consider an operator $\hat A$ defined as a Kronecker product of $I$ and $Z$ operators, each acting on their own qubit.
It is a degenerate operator with half the eigenvalues $+1$ and half $-1$.
Because its matrix form is diagonal, each bitstring we measure corresponds exactly to an eigenstate.
The parity ($\pm1$) of a given bitstring $z$ is precisely the parity of the substring $z'$ which omits any index corresponding to an $I$ operator in $A$.
For example, if $A = I_5 Z_4 Z_3 I_2 Z_1$ and $z=\ket{00101}$, the substring $z'$ leaves off the second and fifth indices: $z'=\ket{011}$. This string has a weight of two, which is an even parity and therefore corresponds to the eigenvalue $+1$.
The expectation value $\expval{A}$ is once again an average of $+1$ and $-1$, weighted by the frequency of bitstrings corresponding to each of the two states.

Last, let us consider a general Pauli word $\sigma_k$.
Half its eigenvalues are again $+1$ and half $-1$, but bitstrings in the computational basis do not correspond exactly to the eigenstates.
We therefore need to diagonalize $\sigma_k$. Let $A_k$ be the Pauli word which replaces all $X$ and $Y$ in $\sigma_k$ by $Z$, and the operator $U_k$ changes the basis so that $\sigma_k = U_k^\dagger A_k U_k$.
Then, for each expectation value we have $\expval{\sigma_k} = \expval{U_k^\dagger A_k U_k}$. This is equivalent to measuring the expectation value $\expval{A_k}$ in a new state $\ket{\psi'} = U_k \ket{\psi}$.
Thus, we may apply at the end of the variational circuit the sequence of gates representing $U_k$, and then apply the methods of the previous paragraph to evaluate $\expval{\sigma_k}$.
One example of $U_k$ could be an operator applying the Hadamard gate $H$ to each qubit corresponding to an $X$ operator in $\sigma_k$, and the sequence of gates $HSZ$ to each qubit corresponding to a $Y$ operator.

\section{Error Analysis and Mitigation}
Quantum error correction, or more often error mitigation is essential for a reliable attainment of computations on a real QPU.\cite{error_prx, error_mitigation, error_natcom} The quantum measurement, an integral element of any algorithm, is by itself probabilistic. In particular, expectation values of an operator are estimated over a large number (M) of discrete measurements. On a noiseless quantum computer, the variance in the expectation value of the Hamiltonian is limited by
\begin{equation}
\langle \epsilon^2 \rangle \le \frac{\overline{E^2}}{M}
\end{equation}
\noindent where $\overline{E^2}$ is the average of the squared energy. It defines an uncertainty and can be resolved to an arbitrary level of precision by increasing the number of measurements.

Importantly, the qubits may accumulate errors either due to the imprecise manipulation or interactions with environment. One of the major sources of errors that we have identified while collecting the data from the quantum processors is the readout error, emerging due to a certain probability that a qubit in a true $|\vb*{0}\rangle$ state is measured as a $|\vb*{1}\rangle$ or vice versa. Repeated measurements of prepared $|\vb*{0}\rangle$ or $|\vb*{1}\rangle$ states reveal transition rates $w_{01}$ and $w_{10}$, defined as the probability that $|\vb*{0}\rangle$ is erroneously measured as $|\vb*{1}\rangle$ or $|\vb*{1}\rangle$ is measured as $|\vb*{0}\rangle$, respectively. Moreover, the application of a particular circuit element may result in an imperfect transformation of the qubit state. The so-called gate noise is typically classified as a separate source of error but for the purpose of this study we have assumed it to be intrinsic to the readout error.

The procedure of error mitigation is based on the computation of the transition rates $w_{01}$ and $w_{10}$ and deriving an appropriate expression to correct the measured expectation values. In order to estimate these rates, we have explicitly prepared the state $|\vb*{0}\rangle$ ($|\vb*{1}\rangle$) 100,000 times and counted how many $|\vb*{1}\rangle$s ($|\vb*{0}\rangle$s) were measured, which determines the probability that a bit flip occurs on a readout for a given computational state of each qubit. The transition rates need to be measured and updated often to ensure that the correction scheme remains effective across the duration of the trials. In fact, they are calculated every time before the optimization step is reached to take into account changes in behavior of a specific qubit. Figure 6 reports the transition rates $w_{01}$ and $w_{10}$ evaluated for each qubit while computing band energies. The transition rates are sampled once per minute across the duration of a 50 minute run. The rates corresponding to a flip from $|\vb*{1}\rangle$ to $|\vb*{0}\rangle$ seem to oscillate with a period of roughly 18 minutes, suggesting that environmental effects indeed modulate the behavior of qubits.

\begin{figure}[ht]
\includegraphics[scale=0.99]{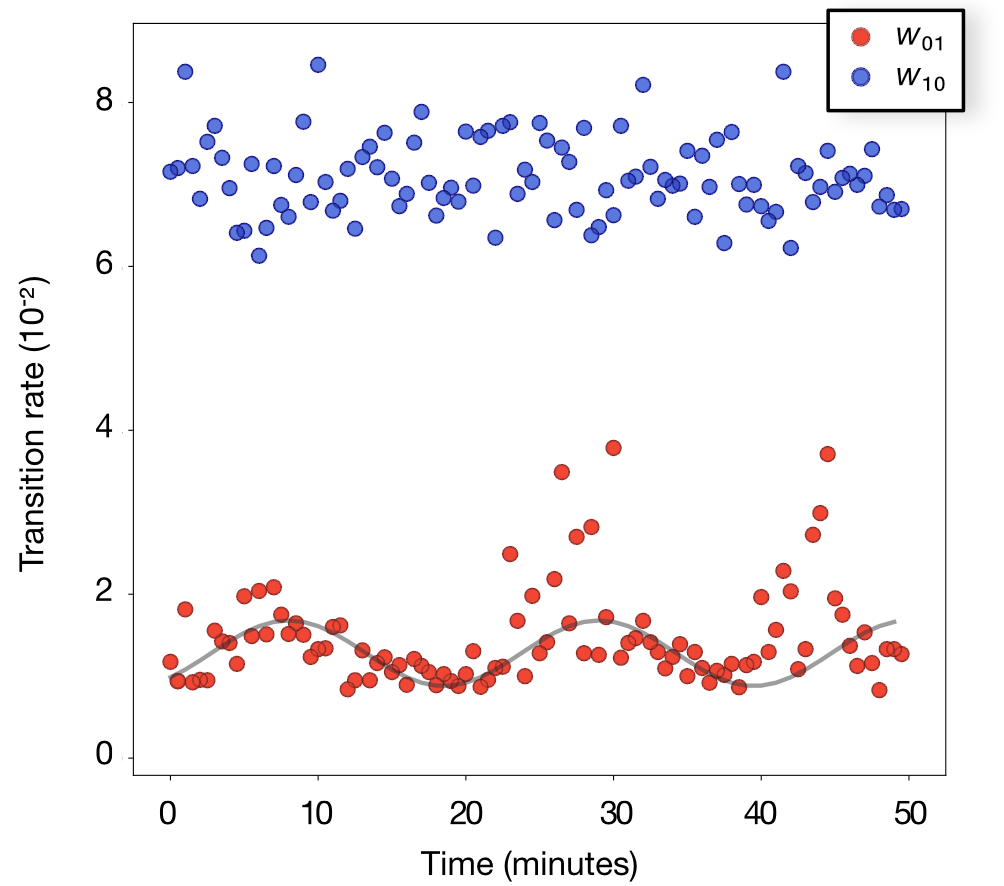}
\caption{Transition rates estimated for a qubit on Rigetti's QPU. Blue circles denote the rates from state $|\vb*{1}\rangle$ to state $|\vb*{0}\rangle$, while red circles correspond to the rates from state $|\vb*{0}\rangle$ to state $|\vb*{1}\rangle$. The fitted trend in transitions suffering from less noise is marked with the gray line. We believe these transitions to arise due the environmental coupling.}
\label{fig:rates}
\end{figure}

The measured expectation value, on a single qubit, can be corrected using the following expression, derived in the Supplementary Material (SM):
\begin{equation}
\langle\hat{\sigma_c}\rangle=\frac{\langle\hat{\sigma}\rangle-p^-}{1-p^+}
\end{equation}
with $p^\pm$ defined in terms of the transition probabilities for the single qubit, $p^\pm=w_{10}\pm w_{01}$. The procedure can be easily generalized to any number of qubits measured in the computational basis,\cite{McCaskey2019} as follows:
\begin{equation}
\langle Z...Z\rangle=\sum_{z\epsilon \mathbb{Z}^n_2}{p(z)\prod^n_{i=1}\frac{(-1)^{z_i}-p^-}{1-p^+}}
\end{equation}
where $z_i$ is the $i^{th}$ element of bitstring $z$, and $z$ is among the set of bitstrings of length $n$ ($\mathbb{Z}^n_2$). The fraction of measured bitstrings resulting in $z$ is denoted as $p(z)$. The correction have been successfully applied to the quantum computation of two-band electronic structure performed on Rigetti. Figure 2(b) shows a comparison between the corrected and uncorrected data points (closed and open circles, respectively), demonstrating that the errors have been significantly reduced.

\section{Summary and perspectives}
In summary, we have computed the band structure of silicon along high symmetry lines in the momentum space using quantum machine accessible via cloud. In order to perform quantum simulations beyond the tractability of modern supercomputers, we need to establish methods of translating a desired physical system to the language of qubits founded with quantum logic gates. The VQE algorithm adapted from quantum chemistry seems to be suitable for electronic structure computation and remarkably, is able to leverage even minimal quantum resources, as demonstrated by the results discussed in this work. In analogy to early quantum chemistry computation tackling the problems with known analytical solutions, we have selected the electronic structure of silicon which is considered trivial in materials science. The presented studies can be thus regarded as a first step towards scalable electronic structure quantum computation that would not be limited to a specific interaction or one particular quantum system. Even though the analyzed Hamiltonian was quite simple, we are convinced that adding interactions, field effects, or corrective terms will be possible in the nearest future.

\begin{acknowledgments}
We thank Rosa Di Felice, Marco Fornari, Ilaria Siloi and Virginia Carnevali for useful discussions. We acknowledge support from the US Department of Energy through the grant \textit{Q4Q: Quantum Computation for Quantum Prediction of Materials and Molecular Properties} (DE-SC0019432). We are also grateful to IBM and Rigetti Computing for providing quantum resources.
\end{acknowledgments}

\bibliography{manuscript}

\end{document}